\documentclass[11pt]{article}
\usepackage{times}
\usepackage{latexsym}
\usepackage{graphicx}
\usepackage{url}
\usepackage{amsmath}
\usepackage{amsthm}
\usepackage{amssymb}

\DeclareMathOperator{\tr}{{\rm Tr}}

\newcommand{\QMA}{\ensuremath{\mathrm{QMA}}}
\newcommand{\StoqMA}{\ensuremath{\mathrm{StoqMA}}}
\newcommand{\MA}{\ensuremath{\mathrm{MA}}}

\newcommand{\AM}{\ensuremath{\mathrm{AM}}}
\newcommand{\NP}{\ensuremath{\mathrm{NP}}}
\newcommand{\BPP}{\ensuremath{\mathrm{BPP}}}
\newcommand{\SBP}{\ensuremath{\mathrm{SBP}}}
\newcommand{\BPPpath}{\ensuremath{\mathrm{BPP_{path}}}}
\newcommand{\POSTBPP}{\ensuremath{\mathrm{PostBPP}}}

\renewcommand{\P}{\ensuremath{\mathrm{P}}}

\newcommand{\be}{\begin{equation}}
\newcommand{\ee}{\end{equation}}
\newcommand{\ba}{\begin{array}}
\newcommand{\ea}{\end{array}}
\newcommand{\bea}{\begin{eqnarray}}
\newcommand{\eea}{\end{eqnarray}}
\newcommand{\calH}{{\cal H }}

\newcommand{\calS}{{\cal S }}

\newcommand{\CC}{\mathbb{C}}

\newcommand{\la}{\langle}
\newcommand{\ra}{\rangle}

\newcommand{\prob}{{\bf Pr}}
\newcommand{\nn}{\nonumber}

\newcommand{\yes}{\mbox{\small \it yes}}
\newcommand{\no}{\mbox{\small \it no}}
\newcommand{\slno}{\Rightarrow}

\newcommand{\ket}[1]{\big|#1\big>}

\newtheorem{theo}{Theorem}

\newtheorem{dfn}{Definition}
\newtheorem{lemma}{Lemma}
\newtheorem{theorem}{Theorem}
\newtheorem{prop}{Proposition}
\newtheorem{cor}{Corollary}

\begin{document}
\title{Merlin-Arthur Games and Stoquastic Complexity}
\author{Sergey Bravyi \thanks{IBM Watson Research Center, P.O. Box 218, Yorktown Heights, NY, USA 10598.
\texttt{sbravyi@us.ibm.com}} \and Arvid J. Bessen \thanks{Columbia
University, New York, NY USA. \texttt{bessen@cs.columbia.edu}}\and
 \and Barbara M. Terhal \thanks{IBM
Watson Research Center, P.O. Box 218, Yorktown Heights, NY, USA
10598. \texttt{bterhal@gmail.com}}}

\maketitle
\setcounter{page}{0}
\begin{abstract}
\MA{} is a class of decision problems for which `yes'-instances have
a proof that can be efficiently checked by a classical randomized
algorithm. We prove that \MA{} has a natural complete problem which
we call the stoquastic $k$-SAT problem. This is a matrix-valued
analogue of the satisfiability problem in which clauses are
$k$-qubit projectors with non-negative matrix elements, while a
satisfying assignment is a vector that belongs to the space spanned
by these projectors. Stoquastic $k$-SAT is the first non-trivial
example of a \MA-complete problem. We also study the minimum
eigenvalue problem for local stoquastic Hamiltonians that was
introduced in Ref. \cite{BDOT06}, stoquastic LH-MIN. A new
complexity class \StoqMA{} is introduced so that stoquastic LH-MIN
is \StoqMA-complete. We show that $\MA{} \subseteq \StoqMA{}
\subseteq {\rm SBP} \cap {\rm QMA}$. Lastly, we consider the average
LH-MIN problem for local stoquastic Hamiltonians that depend on a
random or `quenched disorder' parameter, stoquastic AV-LH-MIN. We
prove that stoquastic AV-LH-MIN is contained in the complexity class
\AM{}, the class of decision problems for which yes-instances have a
randomized interactive proof with two-way communication between
prover and verifier.
\end{abstract}

\newpage
\setcounter{page}{1}
\section{Introduction}

Recent years have seen the first steps in the development of a
quantum or matrix-valued complexity theory. Such complexity theory
is interesting for a variety of reasons. Firstly, as in the
classical case it may increase our understanding of the power and
limitations of quantum computation. Secondly, since quantum
computation is an extension of classical computation, this
complexity theory provides a framework and new angle from which we
can view classical computation.

In this paper we will provide such a new point of view for the
complexity class \MA{} defined by Babai~\cite{babai}.
 We do this by studying so-called stoquastic problems, first
defined in \cite{BDOT06}. The first problem we consider is one that
arises naturally through a quantum or matrix-valued generalization
of the satisfiability problem \cite{Bravyi06}. The input of quantum
$k$-SAT is a tuple $(n,\epsilon,\Pi_1,\ldots,\Pi_M,S_1,\ldots,S_M)$,
where $n$ is a number of qubits, $\epsilon\ge n^{-O(1)}$ is a
precision parameter, and $\Pi_1,\ldots,\Pi_M$ are Hermitian
projectors acting on the Hilbert space of $n$ qubits. Each projector
$\Pi_\alpha$ acts non-trivially only on some subset of $k$ qubits
$S_\alpha \subseteq \{1,2,\ldots,n\}$. Then the promise problem
quantum $k$-SAT is stated as follows:
\begin{itemize}
\item {\it yes-instance}: There exists a state $|\theta\ra\in (\CC^2)^{\otimes n}$ such that
for all $\alpha=1,\ldots,M$,  $\Pi_\alpha\, |\theta\ra=|\theta\ra$.
\item {\it no-instance}: For any state $|\theta\ra\in (\CC^2)^{\otimes n}$ there is some
$\alpha\in \{1,\ldots,M\}$ such that $\la \theta |\Pi_\alpha
|\theta\ra \le 1-\epsilon$.
\end{itemize}
(Here a {\it state} is a vector $|\theta\ra\in (\CC^2)^{\otimes n}$
with a unit norm $\la \theta |\theta\ra =1$.) A state $|\theta\ra$
satisfying the condition for a yes-instance is called a solution, or
a satisfying assignment.

If the projectors $\Pi_\alpha$ have zero off-diagonal elements in
the computational basis, a solution $|\theta\ra$ can always be
chosen as a basis vector, $|\theta\ra=|x\ra$, $x\in \{0,1\}^n$. In
this case quantum $k$-SAT reduces to classical $k$-SAT which is
known to be \NP-complete for $k\ge 3$. On the other hand, if no
restrictions on the matrix elements of $\Pi_\alpha$ are imposed,
quantum $k$-SAT is complete for Quantum \MA, or \QMA, defined by
Kitaev~\cite{KSV:computation,QMA4} if $k\ge 4$, see~\cite{Bravyi06}.
The class \QMA{} has been extensively studied
in~\cite{QMA1,QMA2,QMA3,QMA4,QMA5,QMA6,QMA7,QMA8,QMA9}. It was
proved that quantum $2$-SAT has an efficient classical
algorithm~\cite{Bravyi06} similar to classical $2$-SAT.

Let us now properly define the restriction that defines the
stoquastic $k$-SAT problem:
\begin{dfn}
Stoquastic $k$-SAT is defined as quantum $k$-SAT with the
restriction that all projectors $\Pi_\alpha$ have real non-negative
matrix elements in the computational basis.
\end{dfn}
The term `stoquastic' was introduced in Ref. \cite{BDOT06} to
suggest the relation both with stochastic processes and quantum
operators. We will show that
\begin{theorem}\label{theo:ma}
Stoquastic $k$-SAT is contained in \MA{} for any constant $k$ and
\MA-hard for $k\ge 6$.
\end{theorem}
It follows that stoquastic $6$-SAT is \MA-complete. This is the
first known example of a natural \MA-complete problem. The proof of
the theorem involves a novel polynomial-time random-walk-type
algorithm that takes as input an instance of stoquastic $k$-SAT and
a binary string $x\in \{0,1\}^n$. The algorithm checks whether there
exists a solution $|\theta\ra$ having large enough overlap with the
basis vector $|x\ra$. Description of such a basis vector can serve
as a proof that a solution exists. The proof of
Theorem~\ref{theo:ma} is given in Section~\ref{sec:inMA}.

Our second result concerns the complexity class \AM{} (Arthur-Merlin
games). \AM{} is a class of decision problems for which
`yes'-instances have a randomized interactive  proof with a constant
number of communication rounds between verifier Arthur and prover
Merlin. By definition, $\MA{} \subseteq \AM$. It was shown that
\AM{} contains some group theoretic problems~\cite{babai}, the graph
non-isomorphism problem~\cite{GMW91} and the approximate set size
problem \cite{GS:coins}.
We show that there exists an interesting quantum mechanical problem
that is in \AM{} (and in fact \AM{}-complete). It is closely related
to the minimum eigenvalue problem for a local
Hamiltonian~\cite{KSV:computation} which we shall abbreviate as
LH-MIN. The input of LH-MIN is a tuple
$(n,H_1,\ldots,H_M,S_1,\ldots,S_M,\lambda_{\yes}, \lambda_{\no})$,
where $n$ is the total number of qubits, $H_\alpha$ is a Hermitian
operator on $n$ qubits acting non-trivially only on a subset of $k$
qubits $S_\alpha \subseteq \{1,2,\ldots,n\}$, and
$\lambda_{\yes}<\lambda_{\no}$ are real numbers. It is required that
$||H_\alpha||\le n^{O(1)}$ and $\lambda_{\no}-\lambda_{\yes}\ge
n^{-O(1)}$. The promise problem LH-MIN is stated as follows:
\begin{itemize}
\item {\it yes-instance}: There exists a state $|\theta\ra\in (\CC^2)^{\otimes n}$ such that
$\sum_{\alpha=1}^M \la \theta|H_\alpha|\theta\ra \le
\lambda_{\yes}$.
\item {\it no-instance}: For any state $|\theta\ra\in (\CC^2)^{\otimes n}$  one has
$\sum_{\alpha=1}^M \la \theta |H_\alpha |\theta\ra \ge
\lambda_{\no}$.
\end{itemize}
In other words, the minimum eigenvalue $\lambda(H)$ of a $k$-local
Hamiltonian $H=\sum_\alpha H_\alpha$ obeys  $\lambda(H)\le
\lambda_{\yes}$ for yes-instances and $\lambda(H)\ge \lambda_{\no}$
for no-instances.

LH-MIN for 2-local Hamiltonians can be viewed as the natural
matrix-valued generalization of MAX2SAT which is the problem of
determining the maximum number of satisfied clauses where each
clause has two variables. It was shown
in~\cite{KSV:computation,QMA4} that LH-MIN is \QMA-complete for
$k\ge 2$. The authors in Ref.~\cite{BDOT06} considered the LH-MIN
problem for so-called stoquastic Hamiltonians.
\begin{dfn}
Stoquastic {\rm LH-MIN} is defined as {\rm LH-MIN} with the
restriction that all operators $H_\alpha$ have real non-positive
off-diagonal matrix elements in the computational basis.
\end{dfn}
 The important
consequence of this restriction is that the eigenvector with lowest
eigenvalue, also called the {\em ground-state}, of a
Hamiltonian $H=\sum_\alpha H_\alpha$ is a vector with nonnegative coefficients in the
computational basis. This allows for an interpretation of this
vector as a probability distribution. For a general Hamiltonian the
ground-state is a vector with complex coefficients for which no such
representation exists. Besides, stoquastic $k$-SAT is a special case
of $k$-local stoquastic  LH-MIN (choose $\Pi_\alpha$ as a projector
onto the space on which $H_\alpha$ takes its smallest eigenvalue
$\lambda_\alpha$ and choose $\lambda_{\yes} = \sum_\alpha
\lambda_\alpha$). The authors in Ref.~\cite{BDOT06} have proved that
 (i) the complexity of stoquastic LH-MIN does not depend on the locality
parameter $k$ if $k\ge 2$; (ii) stoquastic LH-MIN is hard for \MA;
(iii) stoquastic LH-MIN is contained in any of the complexity
classes \QMA, \AM, \POSTBPP{} (the latter inclusion was proved only
for Hamiltonians with polynomial spectral gap), where
\POSTBPP=\BPPpath, see~\cite{BDOT06,Han:threshold}.

In the present paper we formulate a random stoquastic LH-MIN problem
that we prove to be complete for the class \AM. In fact the most
interesting aspect of this result is that this problem is contained
in \AM{}, since it is not hard to formulate a complete problem for \AM,
 see below.
 Let us define this problem stoquastic AV-LH-MIN properly.
We consider an ensemble of local stoquastic Hamiltonians $\{H(r)\}$
for which $r$ is a string of $m=n^{O(1)}$ bits, and $r$ is taken
from the uniform distribution on $\Sigma^m$. Such a random ensemble
$\{H(r)\}$ is called $(k,l)$-local if $H(r)$ can be written as
$H(r)=\sum_{\alpha=1}^M H_{\alpha}(r), \quad M=n^{O(1)}$, where
$H_{\alpha}(r)$ is a Hermitian operator on $n$ qubits acting
non-trivially only on some subset of qubits $S_\alpha\subseteq
\{1,\ldots,n\}$, $|S_\alpha|\le k$. Furthermore, $H_{\alpha}(r)$
depends only on some subset of random bits $R_\alpha\subseteq
\{1,\ldots,m\}$, $|R_\alpha|\le l$. We will consider ensembles in
which the Hamiltonians $H(r)$ are stoquastic, i.e. each
$H_{\alpha}(r)$ has real non-positive off-diagonal matrix elements
for all $r$ \footnote{Note that this property can be efficiently
verified since we have to test only $2^l$ random bit
configurations.}. The input of the problem stoquastic AV-LH-MIN
involves a description of a $(k,l)$-local stoquastic ensemble
$\{H(r)\}$ on $n$ qubits and $m$ random bits, and two thresholds
$\lambda_{\yes}<\lambda_{\no}$. It is required that
$||H_\alpha(r)||\le n^{O(1)}$ for all $r$, and
$\lambda_{\no}-\lambda_{\yes}\ge n^{-O(1)}$. Let us denote by
$\lambda(r)$ the smallest eigenvalue of $H(r)$ and
$\bar{\lambda}=2^{-m} \sum_{r\in \Sigma^m} \lambda(r)$ the average
value of $\lambda(r)$. The stoquastic AV-LH-MIN problem is to decide
whether $\bar{\lambda}\le \lambda_{\yes}$ (a yes-instance) or
$\bar{\lambda}\ge \lambda_{\no}$ (a no-instance). Our second result
is
\begin{theorem}
\label{theo:am}
Stoquastic {\rm AV-LH-MIN} is contained in \AM{} for any
$k,l=O(1)$. Stoquastic $(3,1)$-local \mbox{AV-LH-MIN} is AM-complete. \label{theo:am}
\end{theorem}

The proof of the theorem is presented in Section~\ref{sec:am}.
It should be mentioned that the stoquastic $(3,1)$-local ensemble $\{H(r)\}$  corresponding
to \AM-hard problem in Theorem~\ref{theo:am} is actually an ensemble of classical $3$-SAT problems,
that is, for each random string $r$ all operators $H_\alpha(r)$ in the decomposition
$H(r)=\sum_\alpha H_\alpha(r)$ are projectors diagonal in the computational basis.
For yes-instance of the problem one has $\lambda(r)=0$ for all $r$ (and thus $\bar{\lambda}=0$), while
for no-instances $\lambda(r)= 0$ with probability at most $1/3$ (and thus $\bar{\lambda}\ge 2/3$),
see Section~\ref{sec:am}. Since classical $3$-SAT is a special case of stoquastic $3$-SAT,
we conclude that a $(3,1)$-local ensemble of stoquastic $3$-SAT problems also yields an \AM-complete
problem.

Our final result concerns the complexity of stoquastic LH-MIN
(without disorder). We define a new complexity class \StoqMA{} which
sits between \MA{} and \QMA{} and we prove, see Section
\ref{sec:stoqma}, that

\begin{theorem}
Stoquastic $k$-local {\rm LH-MIN} is \StoqMA{}-complete for any $k\ge 2$.
\label{theo:stoqma}
\end{theorem}

\begin{figure}[h]
\centerline{ \mbox{
 \includegraphics[height=4cm]{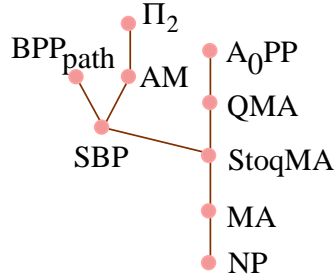}
 }}
\caption{ \label{fig:diagram} Inclusion tree for the relevant
complexity classes. Here \BPPpath=\POSTBPP.}
\end{figure}

The class \StoqMA{} is a restricted version of \QMA{} in which the
verifier can perform only classical reversible gates, prepare qubits
in $|0\ra$ and $|+\ra$ states, and perform one measurement in the
$|+\ra,|-\ra$ basis. This results solves the open problem posed
in~\cite{BDOT06} concerning the complexity of stoquastic LH-MIN. We
also establish some relations between \StoqMA{} and already known
complexity classes. Ref.~\cite{boe-gla-mei-03} introduced a
complexity class \SBP{} (Small Bounded-Error Probability) as a
natural class sitting between \MA{} and \AM{}. We prove that
stoquastic LH-MIN and thus all of \StoqMA{} is contained in \SBP,
see Section~\ref{sec:insbp} for details. Figure~\ref{fig:diagram}
illustrates the relevant complexity classes and their
inter-relations.

In conclusion, our results show that the randomized versions of
stoquastic LH-MIN, stoquastic $k$-SAT and classical $k$-SAT are
of equal complexity, that is they are all \AM{}-complete. On the
other hand, it is at present unclear whether the original problems
(not randomized) $k$-SAT, stoquastic
$k$-SAT and stoquastic LH-MIN  and thus the corresponding classes
\NP{}, \MA{} and \StoqMA{} are of equal complexity. We would like to
note that any proof of a separation between \MA{} and \AM{} (for
example via a separation of \MA{} and \StoqMA{}) would have
far-reaching consequences. Namely it was proved in \cite{AKSS:MAAM}
that

\begin{theorem}[\cite{AKSS:MAAM}]
If $\MA{} \neq \AM{}$ then $\NP{} \not \subseteq \P{}/{\rm poly}$.
\end{theorem}

\section{Definitions of relevant complexity classes}
\label{sec:cc} Throughout the paper $\Sigma^n$ and $\Sigma^*$ will
denote a set of $n$-bit strings and the set of all finite bit strings
respectively.


\begin{dfn}[\MA{}]\label{def:MA}
A promise problem $L=L_{\yes}\cup L_{\no}\subseteq \Sigma^*$ belongs
to \MA{} iff there exist a polynomial $p(n)$ and a \BPP{} predicate
$V(x,w)$
such that
\bea
x\in L_{\yes} &\slno & \exists \, w \quad
\prob{\left[V(x,w)=1\right]}\ge 2/3\;\;  \mbox{\rm (Completeness)} \label{comp} \nn  \\
x\in L_{\no} & \slno & \forall\, w \quad
\prob{\left[V(x,w)=1\right]}\le 1/3 \;\; \mbox{\rm (Soundness)} \nn
\label{sound}
 \eea
Here $x\in \Sigma^*$ represents the
 instance of a problem and $w\in \Sigma^{p(|x|)}$ represents the prover's witness string.
 If an instance $x$ does not satisfy the promise,
i.e., $x\notin L_{\yes}\cup L_{\no}$, then $V(x,w)$ may be arbitrary
(or even undefined).
\end{dfn}

In \cite{BDOT06} it was proved that \MA{} has an alternative
quantum-mechanical definition as a restricted version of \QMA{} in
which the verifier is a coherent classical computer, see the review
in Section \ref{subs:ccv} of the Appendix. \StoqMA{} is a class of
decision problems for which the answer `yes' has a short quantum
certificate that can be efficiently checked by a {\it stoquastic
verifier}:

\begin{dfn}[\StoqMA{}]
A stoquastic verifier is a tuple $V=(n,n_w,n_0,n_+,U)$, where $n$ is
the number of input bits, $n_w$ the number of input witness qubits,
$n_0$ the number of input ancillas $|0\ra$, $n_+$ the number of
input ancillas $|+\ra$ and $U$ is a quantum circuit on
$n+n_w+n_0+n_+$ qubits with $X$, CNOT, and Toffoli gates. The
acceptance probability of a stoquastic verifier $V$ on input string
$x\in \Sigma^n$ and witness state $|\psi\ra\in (\CC^2)^{\otimes
n_w}$ is defined as $\prob(V;x,\psi)=\la \psi_{in}|U^\dag\, \Pi_{out}\, U
|\psi_{in}\ra$. Here $|\psi_{in}\ra = |x\ra\otimes |\psi\ra\otimes |0\ra^{\otimes
n_0}\otimes |+\ra^{\otimes n_+}$ is the initial state and
$\Pi_{out}=|+\ra\la +|_1\otimes I_{else}$ projects the first qubit
onto the state $|+\ra$.\\
A promise problem $L=L_{\yes}\cup
L_{\no}\subseteq \Sigma^*$ belongs to \StoqMA{} iff there exists a
uniform family of stoquastic verifiers, such that for any fixed
number of input bits $n$ the corresponding verifier $V$ uses at most
$n^{O(1)}$ qubits, $n^{O(1)}$ gates, and obeys completeness and
soundness conditions: \bea  \quad x\in
L_{\yes} &\slno & \exists \, |\psi\ra\in (\CC^2)^{\otimes n_w}
  \quad
\prob(V;x,\psi)\ge \epsilon_{\yes}\;\; \mbox{\rm (Completeness)} \nn \\
\quad x\in L_{\no} &\slno & \forall \, |\psi\ra
\in (\CC^2)^{\otimes n_w} \quad  \prob(V;x,\psi)\le
\epsilon_{\no}\;\;\mbox{\rm (Soundness)}\nn\label{sound1}
\end{eqnarray}
Here the threshold probabilities $0\le \epsilon_{\no}<\epsilon_{\yes}\le 1$
must have polynomial separation: $\epsilon_{\yes}-\epsilon_{\no}\ge n^{-O(1)}$.
\label{dfn:StoqMA}
\end{dfn}

\noindent {\it Comments:} In contrast to the standard classes \BPP,
\MA,  or \QMA{} the class \StoqMA{} does not permit amplification of
the gap between the threshold probabilities $\epsilon_{\no}$,
$\epsilon_{\yes}$ based on majority voting. In fact, it is not hard
to show that the state $\psi$ maximizing the
acceptance probability has non-negative amplitudes in the computational
basis, and $\prob(V;x,\psi) \in
[\frac{1}{2},1]$ for any non-negative state $|\psi\ra$.

It is important to note that the only difference between \StoqMA{}
and \MA{} is that a stoquastic verifier in \StoqMA{} is allowed to
do the final measurement in the $\{|+\ra,|-\ra\}$ basis, whereas a
classical coherent verifier in \MA{} can only do a measurement in
the standard basis $\{\ket{0},\ket{1}\}$.



The complexity class \AM{} was introduced by
Babai~\cite{babai} as a class of decision problems for which the answer `yes'
possesses a randomized interactive proof (Arthur-Merlin game) with two-way communication between
a prover and a verifier. Babai also showed in~\cite{babai} that any
language in \AM{} has a proving protocol such that (i) verifier sends prover a uniform random bit string $q$;
(ii) prover replies with a witness string $w$; (iii) verifier
performs polynomial-time deterministic computation on $q$ and $w$ to decide whether he
accepts the proof.
Here is a formal definition:

\begin{dfn}[\AM{}]
A promise problem  $L=L_{\yes}\cup L_{\no}\subseteq \Sigma^*$
belongs to the class \AM{} iff there exists a polynomial $p$ and a
\P{} predicate $V(x,q,w)$ defined for any $q,w\in \Sigma^{p(|x|)}$,
such that \bea x\in L_{\yes} &\slno &
\prob{\left[\exists \, w \, : \, V(x,q,w)=1\right]}\ge 2/3 \;\; \mbox{\rm (Completeness)}\nn \\
x\in L_{\no} &\slno &
\prob{\left[\exists \, w \, : \, V(x,q,w)=1\right]}\le 1/3 \;\; \mbox{\rm (Soundness)} \nn\label{AM}
\eea
where $q\in \Sigma^{p(|x|)}$ is a uniformly distributed random bit string.
\label{def:am}
\end{dfn}

Finally, the complexity class \SBP{} (Small Bounded-error
Probability) was introduced in~\cite{boe-gla-mei-03} as a natural
class sitting between \MA{} and \AM.

\begin{dfn}[\SBP{}]\label{dfn:SBP}
A promise problem  $L=L_{\yes}\cup L_{\no}\subseteq \Sigma^*$ belongs to the class \SBP{}
iff there exists a function $F\in \# $P and
a function $f\, :\,  \Sigma^* \to \mathbb{R}_+$
computable in polynomial time such that
\bea
x\in L_{\yes} &\slno & F(x)\ge f(x)  \quad \mbox{\rm (Completeness)}\nn \\
x\in L_{\no} &\slno & F(x)\le (1/2)\, f(x) \quad \mbox{\rm (Soundness)}\nn
\eea
\end{dfn}
 It was
proved in~\cite{boe-gla-mei-03} that $\SBP{} \subseteq \AM{} \cap \BPPpath$, where
$\BPPpath=\POSTBPP$, see~\cite{BDOT06}.

\section{Stoquastic $6$-SAT is \MA-complete}
\label{subs:macomplete}

We first argue that stoquastic $k$-SAT is \MA-hard for any $k\ge 6$.
This result is a simple corollary of Lemma~3 in Ref.~\cite{BDOT06}
which showed that LH-MIN for a 6-local stoquastic Hamiltonian is
\MA{}-hard (a more formal proof of this result is also given in
Appendix~\ref{subs:ccv}). Indeed, let $L=L_{\yes}\cup L_{\no}$ be
any language in \MA{} and let $V$ be a verifier for this language,
see Definition~\ref{def:MA}. Without loss of generality $V$ accepts
with probability $1$ on `yes'-instances,
see~\cite{furer89completeness}. As was shown in Ref.~\cite{BDOT06},
for every input $x\in L$ one can construct a stoquastic $6$-local
Hamiltonian $H=\sum_{\alpha} H_\alpha$ such that
$\lambda(H)=\lambda_{\yes}=\sum_{\alpha} \lambda(H_\alpha)$ for
$x\in L_{\yes}$ and $\lambda(H)\ge \lambda_{\no}=\lambda_{\yes} +
|x|^{-O(1)}$ for $x\in L_{\no}$. The corresponding LH-MIN problem is
thus equivalent to quantum $6$-SAT with projectors $\Pi_\alpha$
projecting onto the ground-space of $H_\alpha$. Such a projector has
non-negative matrix elements because $H_\alpha$ has non-positive
off-diagonal matrix elements. Therefore any problem in \MA{} can be
reduced to stoquastic $6$-SAT:
\begin{cor}
Stoquastic $6$-{\rm SAT} is \MA-hard. \label{cor:mahard}
\end{cor}

\subsection{Stoquastic $k$-{\rm SAT} is contained in \MA}
\label{sec:inMA}

In this section we describe a random-walk-type algorithm for
stoquastic $k$-SAT. Given an instance of stoquastic $k$-SAT with the
projectors $\{\Pi_{\alpha}\}$ we can define a Hermitian operator \be
\label{G} G=\frac1{M} \sum_{\alpha=1}^M \Pi_\alpha. \ee Note also
that $G$ has non-negative matrix elements in the computational
basis. We have that either the largest eigenvalue of $G$ is
$\lambda=1$ (a yes-instance) or $\lambda\le 1-\epsilon M^{-1}$ (a
no-instance) since for any vector $|\theta\ra$
\[
\la \theta|G|\theta\ra \le 1-M^{-1} + M^{-1}\min_\alpha \la
\theta|\Pi_\alpha |\theta\ra \le 1-\epsilon M^{-1}.
\]

In order to distinguish $\lambda=1$ and $\lambda\le 1-\epsilon
M^{-1}$ the verifier Arthur will employ a random walk on the space
of $n$-bit binary strings. The transition probability from a string
$x$ to a string $y$ will be proportional to the matrix element
$G_{x,y}$. The role of the prover Merlin is to provide the starting
point for the random walk. Each step of the random walk will include
a series of tests that are always passed for positive instances. For
negative instances the tests are passed with probability strictly
less than 1 such that the probability for the random walk to make
$L$ steps decreases exponentially with $L$.

In order to illustrate the main idea, we will first define the walk
for positive instances only. Suppose that a state \be \label{theta}
|\theta\ra = \sum_{x\in T} \theta_x \, |x\ra, \quad \theta_x>0,
\quad T\subseteq \Sigma^n \ee is a satisfying assignment\footnote{We
can always choose a satisfying assignment with non-negative
amplitudes. Indeed, assume $\Pi_\alpha\, |\theta\ra=|\theta\ra$ for
some $|\theta\ra=\sum_x \theta_x \, |x\ra$.
Define $|\tilde{\theta}\ra=\sum_x |\theta_x| \, |x\ra$. Then
$\la \tilde{\theta}|\Pi_\alpha|\tilde{\theta}\ra \ge \la \theta|\Pi_\alpha |\theta\ra=1$
and thus $\Pi_\alpha\,
|\tilde{\theta}\ra= |\tilde{\theta}\ra$.}, that is
$\Pi_\alpha\, |\theta\ra=|\theta\ra$ for all $\alpha$. For any
binary strings $x,y\in T$ define a transition probability from $x$
to $y$ as \be\label{GtoP} P_{x\to y} =  G_{x,y} \left(
\frac{\theta_y}{\theta_x} \right), \quad G_{x,y}=\la x|G|y\ra. \ee
Clearly, $\sum_{y\in T} P_{x\to y}=1$ for all $x\in T$, so that
$P_{x\to y}$ defines a random walk on $T$. A specific feature of
solutions of stoquastic $k$-SAT is that the ratio
$\theta_y/\theta_x$ in Eq.~(\ref{GtoP}) can be easily expressed in
terms of matrix elements of $\Pi_\alpha$, namely one can prove that
\begin{lemma}\label{lemma:key}
Assume $\Pi\, : \, \CC^{2^n} \to \CC^{2^n}$ is a Hermitian projector
having non-negative matrix elements in the computational basis.
Assume $\Pi \, |\theta\ra = |\theta\ra$ for some state
$|\theta\ra=\sum_{x\in T} \theta_x \, |x\ra$, $\theta_x>0$,
$T\subseteq \Sigma^n$.
Then\\ \\
(1) $\la x|\Pi|x\ra >0$ for all $x\in T$,\\ \\
(2) If $\la x|\Pi|y\ra>0$ for some $x,y\in T$ then \be\label{ratio}
\frac{\theta_y}{\theta_x} = \sqrt{\frac{\la y|\Pi|y\ra}{\la
x|\Pi|x\ra}}. \ee
\end{lemma}
The proof of this lemma can be found in
Appendix~\ref{subs:keylemma}. Applying the lemma to Eq.~(\ref{GtoP})
we conclude that either $G_{x,y}=P_{x\to y}=0$ or $G_{x,y}>0$ and
thus \be\label{Px2y} P_{x\to y}=G_{x,y}\, \sqrt{\frac{\la
y|\Pi_\alpha|y\ra}{\la x|\Pi_\alpha|x\ra}} \ee for any $\alpha$ such
that $\la y|\Pi_\alpha |x\ra>0$ (since $G_{x,y}>0$ there must exist
at least one such $\alpha$). Thus for any fixed $x,y \in T$ we can
compute the transition probability $P_{x\to y}$ efficiently. Let,
for any fixed $x\in T$, the set of points $y\in T$ that can be
reached from $x$ by one step of the random walk be $ N(x)=\{ y\in
\Sigma^n\, : \, G_{x,y}>0\}$. This set contains at most $2^k\, M =
n^{O(1)}$ elements which can be found efficiently since $G$ is a sum
of $k$-qubit operators.

Note that definition of transition probabilities Eq.~(\ref{Px2y})
does not explicitly include any information about the solution
$|\theta\ra$. This is exactly the property we are looking for: the
definition of the random walk must be the same for positive and
negative instances. Of course, applying Eq.~(\ref{Px2y}) to negative
instances may produce unnormalized probabilities, such that
$\sum_{y\in N(x)} P_{x\to y}$ is either smaller or greater than $1$.
Checking normalization of the transition probabilities will be
included into the definition of the verifier's protocol as an extra
test. Whenever the verifier observes unnormalized probabilities, he
terminates the random walk and outputs `reject'. The probability of
passing the tests will be related to the largest eigenvalue of $G$.
If the verifier performs sufficiently many steps of the walk and all
the tests are passed, he gains confidence that the largest
eigenvalue of $G$ is $1$. We shall see that the soundness condition
in Eq.~(\ref{sound}) is fulfilled if the verifier accepts after
making $L$ steps of the random walk, where $L$ obeys inequality
\be\label{L}
 2^{\frac{n}2} \left(1-\frac{\epsilon}M \right)^L \le
\frac13.
\ee
Since $\epsilon= n^{-O(1)}$ and the number of clauses
$M$ is at most $M\le {n\choose k} = n^{O(1)}$ one can satisfy this
inequality with a polynomial number of steps, $L=n^{O(1)}$. The only
step in the definition of the random walk above that can not be done
efficiently is choosing the starting point. It requires the  prover's
assistance. For reasons related to the soundness of the proof, the
prover is required to send the verifier a string $x\in T$ with the
largest amplitude $\theta_x$.

A formal description of the prover's strategy is the following. In
case of a {\it yes-instance} the prover chooses a vector
$|\theta\ra\in (\CC^2)^{\otimes n}$ such that $\Pi_\alpha\,
|\theta\ra=|\theta\ra$ for all $\alpha$. Wlog, $|\theta\ra$ has
positive amplitudes on some set $T \subseteq \Sigma^n$,
see Eq.~(\ref{theta}).
The prover sends the verifier a string $w\in T$ such that $\theta_w
\ge \theta_x$ for all $x\in T$. In case of a {\it no-instance} the
prover may send the verifier an arbitrary string $w\in \Sigma^n$.

Here is a formal description of the verifier's strategy:
\begin{center}
\fbox{%
\parbox{15cm}{
{\bf Step~1:} Receive a string $w\in \Sigma^n$  from the prover. Set $x_0=w$.\\
{\bf Step~2:} \parbox[t]{12cm}{Suppose the current state of the walk
is $x_j$. Verify that
           $\la x_j|\Pi_\alpha|x_j\ra>0$ for all $\alpha$. Otherwise reject.}\\
{\bf Step~3:} Find the set $N(x_j)=\{y\in \Sigma^n\, : \, G_{x_j,y}>0\}$.\\
{\bf Step~4:} For every $y\in N(x_j)$ choose any
 $\alpha=\alpha(y)$ such that
$\la y|\Pi_{\alpha(y)}|x_j\ra>0$.\\
{\bf Step~5:} For every $y\in N(x_j)$ compute a number
\be\label{trans} P_{x_j\to y}= G_{x_j,y} \, \sqrt{\frac{\la
y|\Pi_{\alpha(y)}|y\ra}{\la x_j|\Pi_{\alpha(y)}|x_j\ra}}. \ee
{\bf Step~6:} Verify that $\sum_{y\in N(x_j)} P_{x_j\to y}=1$. Otherwise reject.\\
{\bf Step~7:} If $j=L$ goto Step~10.\\
{\bf Step~8:} Generate $x_{j+1}\in N(x_j)$ according to the transition probabilities $P_{x_j\to x_{j+1}}$.\\
{\bf Step~9:} \parbox[t]{12cm}{Compute and store a number
\be\label{rj} r_{j+1} = \frac{P_{x_j\to x_{j+1}}}{G_{x_j,x_{j+1}}}.
\ee
Set $j\to j+1$ and goto Step~2.}\\
{\bf Step~10:} Verify that $\prod_{j=1}^L r_j \le 1$. Otherwise reject.\\
{\bf Step~11:} Accept.}%
} \label{prot:MA}
\end{center}

\vspace{2mm}

Step~4 deserves a comment. It may happen that there are several
$\alpha$'s with the property $\la y|\Pi_\alpha |x_j\ra>0$. Let us
agree that $\alpha(y)$ is the smallest $\alpha$ satisfying this
inequality. In fact, the definition of the transition probabilities
$P_{x_j\to y}$ should not depend on the choice of $\alpha(y)$ for
the yes-instances, see Lemma~\ref{lemma:key}.
Step~8 might be impossible to implement {\it exactly} when only
unbiased random coins are available. This step can be replaced by
generating $x_{j+1}$ from a probability distribution $P'_{x_j\to y}$
which is $\delta$-close in variation distance to $P_{x_j\to y}$ for
some $\delta=n^{-O(1)}$. This is always possible even with unbiased
random coins.

In Appendix \ref{app:soundcompl} we formally prove the completeness
and soundness of this protocol. As a consequence of this and
Corollary \ref{cor:mahard} we obtain Theorem \ref{theo:ma}.

\section{Stoquastic LH-MIN is \StoqMA-complete}
\label{sec:stoqma}

In this section we will prove Theorem \ref{theo:stoqma}.

%
%
First we show that stoquastic LH-MIN is contained in \StoqMA{}. Let
$H$ be a stoquastic $k$-local Hamiltonian acting on $n$ qubits. It
is enough to show that there exist constants $\alpha>0$, $\beta$,
and a stoquastic verifier $V$ with $n_w=n$ witness qubits, such that
\be\label{HV} \prob(V;x,\psi)= \la \psi| \, \left(-\alpha\, H +
\beta\, I\right)\, |\psi\ra \quad \mbox{for all} \quad |\psi\ra\in
(\CC^2)^{\otimes n_w}, \ee where $x$ is a classical description of
$H$. We shall construct a stoquastic verifier that picks up one
local term in $H$ at random and converts this term into an
observable proportional to $|+\ra\la +|$. This is possible for one
particular decomposition of $H$ into local stoquastic terms which we
shall describe now.
\begin{lemma}\label{lemma:Xform}
Let $H$ be $k$-local stoquastic Hamiltonian on $n$ qubits. There
exist constants $\gamma>0$ and $\beta$ such that \be\label{Xform}
\gamma\, H + \beta\, I = \sum_{j} p_j \, U_j \, H_j \, U_j^\dag, \ee
where $p_j\ge 0$, $\sum_j p_j=1$, $U_j$ is a quantum circuit on $n$
qubits with $X$ and CNOT gates. The stoquastic term $H_j$ is either
$- |0\ra\la 0|^{\otimes k}$ or $- X\otimes |0\ra\la 0|^{\otimes
k-1}$. All terms in the decomposition Eq.~(\ref{Xform}) can be found
efficiently.
\end{lemma}

The next step is to reduce a measurement of the observables $|0\ra\la
0|^{\otimes k}$ and $X\otimes|0\ra\la 0|^{\otimes k-1}$ to a
measurement of $X$ only.

\begin{lemma}\label{lemma:t}
An operator $W\, : \, (\CC^2)^{\otimes p} \to (\CC^2)^{\otimes q}$
is called a stoquastic isometry iff
\[
W\, |\psi\ra = U\, |\psi\ra\otimes |0\ra^{\otimes n_0} \otimes
|+\ra^{\otimes n_+} \quad \mbox{for all} \quad |\psi\ra \in
(\CC^2)^{\otimes p}
\]
for some integers $n_0$ and $n_+$, $q=p+n_0+n_+$, and some quantum
circuit $U$ on $q$ qubits with $X$, CNOT, and Toffoli gates. For any
integer $k$ there exist a stoquastic isometry $W$ mapping $k$ qubits
to $2k+1$ qubits such that
\be\label{ab1}
|0\ra\la 0|^{\otimes k} = W^\dag \left(
X\otimes I^{\otimes 2k}\right)  W.
\ee
Also, for any integer $k$ there exist a stoquastic isometry $W$
mapping $k$ qubits to $2k-1$ qubits such that
\be\label{ab2}
X\otimes |0\ra\la 0|^{\otimes k-1} = W^\dag
\left( X\otimes I^{\otimes 2k-2}\right)  W.
\ee
\end{lemma}

The proof of these Lemmas can be found in Appendix \ref{sec:proofslem}.
Combining Lemmas~\ref{lemma:Xform} and \ref{lemma:t} we get
 \be\label{absum}
\gamma \, H + \beta\, I = -\sum_j p_j \, W_j^\dag (X\otimes I_{else})
W_j, \ee where $\{W_j\}$ is a family
of stoquastic isometries. Clearly, for every term in the sum
Eq.~(\ref{absum}) one can construct a stoquastic verifier $V_j$ such
that
\[
\prob(V_j;x,\psi)=\la \psi|W_j^\dag (|+\ra\la +|\otimes
I_{else}) W_j|\psi\ra.
\]
Here $x$ is a classical description of $H$. Taking into account that
$X=2\, |+\ra\la +|-I$, we get
\[
\la \psi|(-(\gamma/2)\, H + (1-\beta)/2\, I) |\psi\ra  = \sum_j p_j\,
\prob(V_j;x,\psi).
\]
It remains to note that the set of stoquastic verifiers is a convex
set. Indeed, let $V'$ and $V''$ be stoquastic verifiers with the
same number of input qubits and witness qubits. Consider a new
verifier $V$ such that
\[
\prob(V;x,\psi) = (1/2)\, \prob(V';x,\psi) + (1/2)\,
\prob(V'';x,\psi).
\]
Using one extra ancilla $|+\ra$ to simulate a random choice of $V'$
or $V''$, and controlled classical circuits one can easily show that
$V$ is also a stoquastic verifier. Thus we have shown how to
construct a stoquastic verifier satisfying Eq.~(\ref{HV}).

\subsection{Stoquastic LH-MIN is \StoqMA-hard and contained in \SBP{}}
\label{subs:StoqMAhard}

In order to prove that stoquastic LH-MIN is hard for \StoqMA, we
could try to modify the \MA-hardness result of stoquastic LH-MIN
obtained in Ref. \cite{BDOT06}. However Kitaev's
circuit-to-Hamiltonian construction requires a large gap between the
acceptance probabilities for yes versus no-instances (which is
achievable in \MA{} or \QMA{} because of amplification) in order for
the corresponding eigenvalues of the Hamiltonian to be sufficiently
separated. In \StoqMA{} we have no amplification which implies that
a modified construction is needed. This modified construction in
which we add the final measurement constraint as a small
perturbation to the circuit Hamiltonian, is introduced in
Appendix~\ref{subs:ccv}. We show there that for any stoquastic
verifier $V$ with $L$ gates and for any precision parameter $\delta
\ll 1/L^3$ one can define a stoquastic $6$-local Hamiltonian
$\tilde{H}$, see
Eqs.~(\ref{Hclock},\ref{perturbation},\ref{tlambda}), such that its
smallest eigenvalue $\lambda(\tilde{H})$ equals
\[
\lambda(\tilde{H})=\delta (L+1)^{-1}\, \left(1 -  \max_{\psi} \prob(V;\psi,x)\right) + O(\delta^2).
\]
Neglecting the term $O(\delta^2)$ (since $\delta$ can be chosen arbitrarily small as long
as $\delta=n^{-O(1)}$), we get
\bea
\mbox{yes-instance: } \quad \lambda(\tilde{H}) &\le& \lambda_{\yes} = \delta(1 -\epsilon_{\yes})(L+1)^{-1}  \nn \\
\mbox{no-instance: } \quad \lambda(\tilde{H}) &\ge&
\lambda_{\no}=\delta(1 -\epsilon_{\no})(L+1)^{-1}  \nn \eea Since
$\epsilon_{\yes}-\epsilon_{\no}=n^{-O(1)}$, we conclude that
$\lambda_{\no}-\lambda_{\yes}=n^{-O(1)}$. Thus stoquastic $6$-local
LH-MIN is \StoqMA-hard. It remains to note that the complexity of
stoquastic $k$-local LH-MIN does not depend on $k$ (as long as $k\ge
2$), see~\cite{BDOT06}.

\subsubsection{Containment in \SBP{}}
\label{sec:insbp}

We can prove that stoquastic LH-MIN and thus all of \StoqMA{} is
contained in the class SBP.  Our proof is essentially a
straightforward application of the result in Ref. \cite{BDOT06}
which showed that stoquastic LH-MIN was contained in \AM{}. We will
only sketch the ideas of the proof here. Given a stoquastic local
Hamiltonian $H$ we can define a non-negative matrix $G =
\frac{1}{2}(I - H/p(n))$ for some polynomial $p(n)$ such that
all matrix elements $0 \leq G_{x,y}\leq 1$. If we define $\mu_{\yes} =
\frac{1}{2}(I - \lambda_{\yes}/p(n))$,
$\mu_{\no} = \frac{1}{2}(I - \lambda_{\no}/p(n))$, and denote $\mu(G)$ the largest
eigenvalue of $G$, then for any integer $L$ one has
\be
\ba{rcccl}
\lambda(H)\le \lambda_{\yes} &\slno & \mu(G) \ge \mu_{\yes} &\slno& \tr(G^L)\ge (\mu_{\yes})^L \\  \\
\lambda(H)\ge \lambda_{\no} &\slno & \mu(G) \le \mu_{\no} &\slno &  \tr(G^L) \le 2^n\, (\mu_{\no})^L. \\
\ea \ee
In Ref.~\cite{BDOT06} it was shown that $\tr(G^L)$ can be
written as $\tr(G^L)
    =
    \frac{1}{2^{m L}}
    \sum_{s \in  \Sigma^{(m+n)L}}
    F_G(s)$
where $F_G\, : \, \Sigma^{(m+n)L}\to \Sigma$ is a polynomial-time computable
Boolean function
and $m$ is the number of bits needed to write down
a matrix element of $G$.
Now one can define a \#P function $F(x)$ such that $x$ is a description of $G$
(or, equivalently, of $H$)  and
$F(x)=\sum_{s} F_G(s)$. Accordingly, $F(x)\ge 2^{mL} (\mu_{\yes})^L$ if
$x$ describes a yes-instance of LH-MIN and $F(x)\le 2^{mL} 2^n (\mu_{\no})^L$
if $x$ describes a no-instance.
Choosing sufficiently large $L=n^{O(1)}$ such that $2^n\, (\mu_{\no})^L \le (1/2) \, (\mu_{\yes})^L$
and defining $g(x)=2^{mL} (\mu_{\yes})^L$
one can satisfy the completeness and soundness conditions in Def.~\ref{dfn:SBP}.
This implies that
\begin{theo}
\StoqMA{} $\subseteq$ \SBP.
\label{theo:inspb}
\end{theo}

\section{Stoquastic AV-LH-MIN is \AM-complete}
\label{sec:am}

We will firstly prove that stoquastic AV-LH-MIN is in \AM.
We are given a $(k,l)$-local stoquastic ensemble $\{H(r)\}$, where
$H(r)$ acts on $n$ qubits and depends on $m$ random bits $r\in
\Sigma^m$. We are promised that $\bar{\lambda}\le \lambda_{\yes}$
for positive instances and $\bar{\lambda}\ge \lambda_{\no}$ for
negative instances. The first step is to use many independent
replicas of the ensemble $\{H(r)\}$ to make the standard deviation
of $\lambda(r)$ much smaller than the gap
$\lambda_{\no}-\lambda_{\yes}$. More strictly, let us define a new
$(k,l)$-local stoquastic ensemble $\{H'(r')\}$, where
$r'=(r^{(1)},\ldots,r^{(N)})\in \Sigma^{Nm}$ contains $N$
independent samples of the random string $r$, and
\[
H'(r')=\frac1N \sum_{j=1}^N H^{(j)}(r^{(j)}).
\]
Here the total number of qubits is $nN$ and $H^{(j)}(r^{(j)})$ is
the original Hamiltonian $H(r^{(j)})$ applied to the $j$-th replica
of the original system. Let $\lambda'(r')$ be the smallest
eigenvalue of $H'(r')$, $\bar{\lambda}'$ be the mean value of
$\lambda'(r')$, and $\sigma(\lambda')$ be the standard deviation of
$\lambda'(r')$. Clearly,
\[
\bar{\lambda}'=\bar{\lambda}, \quad \mbox{and} \quad \sigma(\lambda')=\frac{\sigma(\lambda)}{\sqrt{N}},
\]
where $\sigma(\lambda)$ is the standard deviation of $\lambda(r)$.
Since all Hamiltonians $H(r)$ are sums of local terms with norm
bounded by $n^{O(1)}$, we have $\sigma(\lambda)=n^{O(1)}$. Therefore
we can choose $N=n^{O(1)}$ such that, say, $\sigma(\lambda')\le
(1/100)\, (\lambda_{\no}-\lambda_{\yes})$.

Now let us choose $\lambda_{\yes}'=\lambda_{\yes}+10\,
\sigma(\lambda')$ and $\lambda_{\no}'=\lambda_{\no}-10\,
\sigma(\lambda')$. Then we still have $\lambda_{\no}'-
\lambda_{\yes}' \ge n^{-O(1)}$ and Chebyshev's inequality implies
that $\prob{[\lambda'(r')\le \lambda_{\yes}']}\ge 99/100$ for a
yes-instance, whereas $\prob{[\lambda'(r')\ge \lambda_{\no}']}\ge
99/100$ for a no-instance. Now we can use the fact that stoquastic
LH-MIN is contained in \AM, see~\cite{BDOT06}. Namely, in order to
verify that $\bar{\lambda}\le \lambda_{\yes}$ the verifier
chooses a random $r'$ and then directly follows the proving protocol
of~\cite{BDOT06} to determine whether $\lambda'(r')\le
\lambda_{\yes}'$. Since a randomly chosen Hamiltonian $H'(r')$
satisfies the promise for stoquastic LH-MIN with probability at
least $0.99$, it will increase the completeness and soundness errors
of the protocol~\cite{BDOT06} at most by $1/100$, which is enough to
argue that stoquastic AV-LH-MIN belongs to \AM.

It remains to prove that stoquastic $(3,1)$-local AV-LH-MIN is \AM-hard.
Let $L=L_{\yes}\cup L_{\no}$ be any language in \AM.
As was shown by Furer et al~\cite{furer89completeness},
definitions of \AM{} with a constant completeness error
and with zero completeness error are equivalent.
Thus we can assume that the \P-predicate $V(x,q,w)$ from Definition~\ref{def:am}
has the following properties:
$x\in L_{\yes}$ implies $\forall q \, \exists \, w\, :  V(x,q,w)=1$,
while $x\in L_{\no}$ implies $\prob{\left[\exists \, w\, :  V(x,q,w)=1\right]}\le 1/3$.
Using an auxiliary binary string $z$ of length $|x|^{O(1)}$ one
can apply the standard Cook-Levin reduction to construct a $3$-CNF formula $C(x,q,w,z)$ such that
$(\exists\,  w\, : \, V(x,q,w)=1)$ iff
$(\exists\,  w,z\, : \, C(x,q,w,z)=1)$. Moreover, w.l.o.g. we can assume that each clause
in $C$ depends on at most one bit of $q$ (otherwise, add an extra clause to $C$ that
copies a bit of $q$ into an auxiliary bit).
Therefore $x\in L_{\yes}$ implies $\forall q \, \exists\,  w,z\, : \, C(x,q,w,z)=1$,
while $x\in L_{\no}$ implies $\prob{\left[\exists\,  w,z\, : \, C(x,q,w,z)=1\right]}\le 1/3$.
For any fixed strings $x$ and $q$ one can regard $C(x,q,w,z)$ as a $3$-CNF
formula with respect to $w$ and $z$, i.e. $C(x,q,w,z)=C_1(w,z)\wedge \ldots \wedge C_M(w,z)$.
The minimal number of unsatisfied clauses in $C(x,q,w,z)$ can be represented as
the minimal eigenvalue of a classical $3$-local Hamiltonian $H(x,q)$ depending on $x$ and $q$
which acts on the Hilbert space spanned by basis vectors $w$ and $z$, namely
$H(x,q)=\sum_{\alpha=1}^M \sum_{w,z} (\neg \, C_\alpha(w,z)) |w,z\ra\la w,z|$.
Setting $\lambda_{\yes}=0$ and $\lambda_{\no}=1$ we get an instance of $(3,1)$-local
AV-LH-MIN such that $\bar{\lambda}=0$ for $x\in L_{\yes}$ and $\bar{\lambda}\ge 2/3$
for $x\in L_{\no}$.

\section*{Acknowledgements}
We would like to thank Scott Aaronson, David DiVincenzo and Alexei Kitaev for useful comments
and discussions. S.B. and B.T.
acknowledge support by NSA and ARDA through ARO contract number W911NF-04-C-0098.
A.B. acknowledges support by DARPA and NSF.

\appendix

\section{Appendix}

\subsection{Proof of Lemma \ref{lemma:key}} \label{subs:keylemma} Let us start by giving
a simple characterization of non-negative projectors.
\begin{prop}\label{prop1}
Let $\Pi\, : \, \CC^N \to \CC^N$ be Hermitian projector (i.e.
$\Pi^2=\Pi$ and $\Pi^{\dagger}=\Pi$) with non-negative matrix
elements, $\la x| \Pi |y\ra \ge 0$, $1\le x,y\le N$. There exist
$q=\mathrm{Rank}(\Pi)$ states $|\psi_1\ra,\ldots,|\psi_q\ra\in
\CC^N$ such that
\begin{enumerate}
\item $\la x|\psi_j\ra\ge 0$ for all $x$ and $j$,
\item $\la \psi_j|\psi_k\ra=\delta_{j,k}$ for all $j,k$,
\item
$\Pi=\sum_{j=1}^q |\psi_j\ra\la \psi_j|.$
\end{enumerate}
\end{prop}
Note that non-negative states are pairwise orthogonal iff they have
support on non-overlapping subsets of basis vectors. Thus the
proposition says that non-negative Hermitian  projectors are
block-diagonal (up to permutation of basis vectors) with each block
being a projector onto a non-negative pure state.
\begin{proof} For any basis vector $|x\ra$
define a ``connected component" \[ T_x=\{y\, : \,  \la x|\Pi|y\ra
>0\}. \] (Some of the sets $T_x$ may be empty.) For any triple
$x,y,z$ the inequalities $\la x|\Pi |y\ra >0$, $\la y|\Pi |z\ra >0$
imply $\la x|\Pi |z\ra >0$ since \[ \la x|\Pi |z\ra = \la x|\Pi^2
|z\ra =\sum_u \la x|\Pi|u\ra \la u |\Pi|z\ra \ge \la x|\Pi |y\ra \la
y|\Pi |z\ra >0. \] Therefore the property $\la x|\Pi |y\ra>0$
defines a symmetric, transitive relation on the set  of basis
vectors and we have
\begin{itemize}
\item $y\in T_x$ implies $T_y=T_x$,
\item $y\notin T_x$ implies $T_y\cap T_x=\emptyset$.
\end{itemize}
Consider a subspace $\calH(T_x)\subseteq \CC^N$ spanned by the basis
vectors from $T_x$. Clearly $\calH(T_x)$ is $\Pi$-invariant. Thus
$\Pi$ is block diagonal w.r.t. decomposition of the whole Hilbert
space into the direct sum of spaces $\calH(T_x)$
and the orthogonal complement where $\Pi$ is zero.
 Moreover, the
restriction of $\Pi$ onto any non-zero subspace $\calH(T_x)$ is a projector
with strictly positive entries. According to the Perron-Frobenius
theorem, the largest eigenvalue of a Hermitian operator with
positive entries is non-degenerate. Thus each block of $\Pi$ has
rank $1$, since a projector has eigenvalues $0, 1$ only.\end{proof}

Now we can easily prove Lemma~\ref{lemma:key}.
\begin{proof}[Proof of Lemma~\ref{lemma:key}]
The statement~(1) can be
proved by contradiction. Assume $x\in T$ and $\la x|\Pi|x\ra=0$.
Then $\Pi\, |x\ra=0$ and thus $\theta_x = \la x|\theta\ra = \la
x|\Pi|\theta\ra = 0$ which is a contradiction since $\theta_x>0$ for
all $x\in T$. The statement~(2) follows from the proposition above. Consider
a decomposition of $\Pi$ into non-negative pairwise orthogonal one-dimensional
projectors:
\[
\Pi=\sum_{j=1}^{q} |\psi_j\ra\la \psi_j|, \quad
q=\mathrm{Rank}(\Pi).
\]
The condition $\la x|\Pi|y\ra>0$ implies that $x$ and $y$ belong to
the same rank-one block of $\Pi$, that is
\bea \Pi\, |x\ra &=& \la
\psi_j|x\ra \, |\psi_j\ra = \sqrt{\la x|\Pi|x\ra}\,
|\psi_j\ra \nn \\
\Pi\, |y\ra &=& \la \psi_j|y\ra \, |\psi_j\ra =
 \sqrt{\la y|\Pi|y\ra}\,
|\psi_j\ra \nn \eea for some block $j$. Now we have
\bea \theta_x
&=& \la x|\theta\ra =\la x|\Pi|\theta\ra =\sqrt{\la x|\Pi|x\ra}\,
\la \psi_j|\theta\ra \nn \\
\theta_y &=& \la y|\theta\ra =\la y|\Pi|\theta\ra =\sqrt{\la
y|\Pi|y\ra}\, \la \psi_j|\theta\ra \nn \eea
Both $\theta_x,\theta_y$
are positive since we assumed $x,y\in T$, so
\[
\frac{\theta_y}{\theta_x} = \sqrt{ \frac{\la y|\Pi|y\ra}{\la
x|\Pi|x\ra}}.
\]
\end{proof}


\subsection{Completeness and Soundness of the MA-verifier Protocol}
\label{app:soundcompl}

\subsubsection{Completeness} \label{subs:complete}
Consider a yes-instance with a satisfying assignment $|\theta\ra$,
see Eq.~(\ref{theta}). We assume that the prover is honest, so that
the verifier receives a string $w\in T$ with the largest amplitude
$\theta_w \ge \theta_x$ for all $x\in T$. We will prove that the
verifier will make $L$ steps of the random walk passing all the
tests with probability $1$.

Indeed, suppose that the current state of the walk is $x_j \in T$.
The test at Step~2 will be passed because of Lemma~\ref{lemma:key},
part~(1). Step~3 is well-defined because the set $N(x_j)$ is
non-empty ($x_j$ itself belongs to $N(x_j)$ since Step~2 implies
$G_{x_j,x_j}>0$), the size of $N(x_j)$ is at most $M 2^k=n^{O(1)}$
and all elements of $N(x_j)$ can be found efficiently. Besides we
have the inclusion $N(x_j)\subseteq T$. Indeed, for any $y\in
N(x_j)$ one has
\[
\theta_y = \la y|\theta\ra = \la y|G|\theta\ra = \sum_{z\in T} \la
y|G|z\ra \la z|\theta\ra \ge G_{y,x_j} \theta_{x_j} >0,
\]
since $x_j\in T$, $G_{y,x_j}>0$, and all matrix elements of $G$ are
non-negative. Therefore $y\in T$. Step~4 is well-defined since
$G_{y,x_j}>0$ implies $\la y|\Pi_\alpha |x_j\ra>0$ for some
$\alpha$. For any $y\in N(x_j)$ the number $P_{x_j\to y}$ in
Eq.~(\ref{trans}) is well-defined since $\la x_j
|\Pi_{\alpha(y)}|x_j\ra>0$, see Step~2. According to
Lemma~\ref{lemma:key}, part~(2), the number $P_{x_j\to y}$ defined
by Eq.~(\ref{trans}) coincides with
\[
P_{x_j\to y} = G_{x_j,y} \left( \frac{\theta_y}{\theta_{x_j}}
\right).
\]
Therefore
\[
\sum_{y\in N(x_j)} P_{x_j\to y}= \sum_{y\in T} P_{x_j\to y} =1.
\]
and the test at Step~6 will be passed. Step~8 is well-defined in the
approximate settings: generate $y\in N(x_j)$ according to
probability distribution $P'_{x_j\to y}$ such that $\| P_{x_j\to y}
- P_{x_j\to y}' \|_1\le \delta$, $\delta=n^{-O(1)}$. Step~9 is
well-defined since $x_{j+1}\in N(x_j)$ and thus $G_{x_j,x_{j+1}}>0$.
Summarizing, the random walk will make $L$ steps with probability
$1$.

As for the last test at Step~10, note that
\[
\prod_{j=1}^L r_j= \left( \frac{\theta_{x_L}}{\theta_{x_0}}
\right)=\left( \frac{\theta_{x_L}}{\theta_{w}} \right) ,
\]
see Lemma~\ref{lemma:key}, part~(2). Taking into account that
$\theta_w \geq \theta_x$ for all $x \in T$, one can see that
$\prod_{j=1}^L r_j\le 1$ for all possible $x_L\in T$ and thus
Step~10 will be passed. Thus the verifier always accepts on positive
instances.

\subsubsection{Soundness} \label{subs:sound} Suppose
the protocol is applied to a no-instance. Let us first discuss the
case when Step~8 is implemented exactly. An approximate
implementation will require only a minor modification.

Let us say that a string $x\in \Sigma^n$ is {\it acceptable} iff it
passes the tests at Step~2 and Step~6 of the verifier's protocol. In
other words, $x$ is acceptable  iff
\begin{enumerate}
\item $\la x|\Pi_\alpha |x\ra >0$ for all $\alpha$,
\item $\sum_{y\in N(x)} P_{x\to y}=1$.
\end{enumerate}
Here $N(x)=\{y\in \Sigma^n\, : \, G_{x,y}>0\}$ and $P_{x\to y}$ is
defined by Eq.~(\ref{trans}) with $x_j\equiv x$. Denote
$T_{acc}\subseteq \Sigma^n$ the set of all acceptable strings (it
may happen that $T_{acc}=\emptyset$).

Clearly, the verifier rejects unless the prover's witness $w$ is
acceptable. Thus we can assume that the random walk starts from
$x_0\in T_{acc}$. If the current state $x_j$ of the random walk is
an acceptable string, the probability distribution $P_{x_j\to y}$ on
the set $y\in N(x_j)$ is well-defined. However, in general $N(x_j)$
is not contained in $T_{acc}$, so the random walk can leave the set
$T_{acc}$ with non-zero probability. Clearly, the probability for
the random walk starting from $x_0\in T_{acc}$ to stay in $T_{acc}$
at every step $j=1,2,\ldots, L$ is
\[
\prob\left(\mbox{RW stays in
$T_{acc}$}\right)=\sum_{x_1,\ldots,x_{L}\in T_{acc}} P_{x_0\to x_1}
\, P_{x_1\to x_2} \cdots P_{x_{L-1}\to x_{L}}.
\]
Taking into account Eq.~(\ref{rj}) one gets
\[
\prob\left(\mbox{RW stays in $T_{acc}$}\right) =
\sum_{x_1,\ldots,x_{L}\in T_{acc}} \left(\prod_{j=1}^L r_j\right)
G_{x_0,x_1} \, G_{x_1,x_2} \cdots G_{x_{L-1},x_L}.
\]
At this point we employ the test at Step~10. Indeed, the verifier
accepts iff the random walk stays in $T_{acc}$ at every step
$j=1,\ldots,L$ {\em and} $\prod_{j=1}^L r_j\le 1$. Thus the
probability for the verifier to accept on an input $w=x_0\in
T_{acc}$ can be bounded from above as
\[
\prob\left(\mbox{the verifier accepts on $x_0$}\right) \le
\sum_{x_1,\ldots,x_{L}\in T_{acc}} G_{x_0,x_1} \, G_{x_1,x_2} \cdots
G_{x_{L-1},x_L}.
\]
Taking into account that all matrix elements of $G$ are
non-negative, we get
\[
\prob\left(\mbox{the verifier accepts on $x_0$}\right)
 \le
\sum_{x_1,\ldots,x_L\in \Sigma^n} G_{x_0,x_1}\cdots G_{x_{L-1},x_L}
= 2^{\frac{n}2} \la x_0|G^L|+\ra,
\]
where $|+\ra = 2^{-n/2} \sum_{x\in \Sigma^n} |x\ra$ is the uniform
superposition of all $2^n$ basis vectors. For negative instances the
largest eigenvalue of $G$ is bounded from above by $1-\epsilon/M$
and thus $\la x_0|G^L|+\ra \le (1-\epsilon/M)^L$ and
\[
\prob\left(\mbox{the verifier accepts on $x_0$}\right)\le
2^{\frac{n}2} \left(1-\frac{\epsilon}M\right)^L \le \frac13.
\]

Now suppose that Step~8 is implemented using a probability
distribution $P'_{x_j\to y}$, such that \be \sum_{y\in N(x_j)}
\left| P_{x_j\to y} - P_{x_j\to y}' \right| \le \delta \quad
\mbox{for any} \quad x_j\in T_{acc}. \label{approxP} \ee One can
easily verify that Eq.~(\ref{approxP}) implies
\[
\left| \sum_{x_1,\ldots,x_L\in T_{acc}} P_{x_0\to x_1} \, P_{x_1\to
x_2} \cdots P_{x_{L-1}\to x_L} - P_{x_0\to x_1}' \, P_{x_1\to x_2}'
\cdots P_{x_{L-1}\to x_L}' \right| \le L\delta.
\]
Thus using an approximate probability distribution at Step~8 leads
to corrections of order $L\delta$ to the overall acceptance
probability. Choosing $\delta\ll L^{-1}$ we can get an acceptance
probability smaller than  $1/2$ which can be amplified to $1/3$
using standard majority voting.

\subsection{Coherent classical verifiers, stoquastic verifiers, and circuit Hamiltonians}
\label{subs:ccv} In this section Kitaev's circuit Hamiltonian
construction~\cite{KSV:computation} is applied to stoquastic
verifiers, see Def.~\ref{dfn:StoqMA}. It is the main technical
element of all the hardness results in our paper. Specifically, it
is used in Subsection~\ref{subs:StoqMAhard} to prove that stoquastic
LH-MIN is hard for \StoqMA. Finally, we use a coherent description
of \MA, see~\cite{BDOT06}, to show that stoquastic $6$-SAT is hard
for \MA, see Subsection~\ref{subs:strictproof}.

\subsubsection{Coherent description of \MA}
\begin{dfn}\label{dfn:ccv}
A coherent classical verifier is a tuple $V=(n,n_w,n_0,n_+,U)$, where
\bea
n &=& \mbox{number of input bits},\nn \\
n_w &=& \mbox{number of witness qubits}, \nn \\
n_0 &=& \mbox{number of ancillas $|0\ra$}, \nn \\
n_+ &=& \mbox{number of ancillas $|+\ra$}, \nn\\
U &=& \mbox{quantum circuit on $n+n_w+n_0+n_+$ qubits with $X$,
CNOT, and Toffoli gates}\nn \eea The acceptance probability of a
coherent classical verifier $V$ on an input string $x\in \Sigma^n$ and
witness state $|\psi\ra\in (\CC^2)^{\otimes n_w}$ is defined as
\[
\prob(V;x,\psi)=\la \psi_{in}|U^\dag\, \Pi_{out}\, U
|\psi_{in}\ra,
\]
where $|\psi_{in}\ra = |x\ra\otimes |\psi\ra\otimes |0\ra^{\otimes
n_0}\otimes |+\ra^{\otimes n_+}$ is the initial state and
$\Pi_{out}=|0\ra\la 0|_1\otimes I_{else}$ projects the first qubit
onto the state $|0\ra$.
\end{dfn}
\begin{lemma}[\cite{BDOT06}]
A promise problem $L=L_{\yes}\cup L_{\no}\subseteq \Sigma^*$ belongs
to \MA{} iff there exists a uniform family of coherent classical
verifiers, such that for any fixed number of input bits $n$ the
corresponding verifier $V$ uses at most $n^{O(1)}$ qubits,
$n^{O(1)}$ gates, and obeys completeness and soundness conditions:
\bea
 x\in L_{\yes} &\slno & \exists \,
|\psi\ra\in (\CC^2)^{\otimes n_w}
  \quad
\prob(V;x,\psi)=1 \quad \mbox{(Completeness)} \nn \\
x\in L_{\no} &\slno & \forall \, |\psi\ra
\in (\CC^2)^{\otimes n_w} \quad \prob(V;x,\psi)\le
1/3 \quad \mbox{(Soundness)}.\nn
\end{eqnarray}
\label{lemma:ccv}
\end{lemma}

\subsubsection{The circuit Hamiltonian}
\label{subs:tch}

Let $V=(n,n_w,n_0,n_+,U)$ be a coherent classical verifier or
stoquastic verifier, where the circuit $U$ consists of $L$ gates,
$U=U_L \cdots U_2\, U_1$. Denote $N=n+n_w+n_0+n_+ + L+2$. Define a
linear subspace $\calH\subseteq (\CC^2)^{\otimes N}$ such that \be
\calH=\left\{ |\Phi\ra=\sum_{j=0}^L U_j\cdots U_0\, |x\ra\otimes
|\psi\ra\otimes |0\ra^{\otimes n_0}\otimes |+\ra^{\otimes
n_+}\otimes |1^{j+1} 0^{L-j+1}\ra, \quad |\psi\ra\in
(\CC^2)^{\otimes n_w} \right\},\label{eq:history}  \ee where $x$ is
some fixed input string and $U_0\equiv I$. States from $\calH$
represent computational paths of the verifier's quantum computer
starting from an arbitrary witness state $|\psi\ra$. For any fixed
$|\psi\ra$ all $L+1$ computational states along the path starting
from $|\psi\ra$ are taken in a superposition and `labeled' by
pairwise orthogonal `clock states' $|1^{j+1} 0^{L-j+1}\ra$,
$j=0,\ldots,L$. It is convenient to label the clock qubits by
$j=0,\ldots,L+1$. Note that the clock qubit $0$ is always set to
$1$, while the clock qubit $L+1$ is always set to $0$. For any
$j=1,\ldots,L$,  the clock qubit $j$ is a flag telling whether the
gate $U_j$ has or has not been applied.

Let us show that $\calH$ is spanned by solutions of a
stoquastic $6$-SAT problem. Introduce non-negative $3$-qubit projectors
\[
\Pi^{init\; x}_j =|x_j\ra\la x_j|_{input\; j} \otimes |10\ra\la
10|_{clock\; 0,1} + |11\ra\la 11|_{clock\; 0,1}, \quad j=1,\ldots,n,
\]
\[
\Pi^{init\; 0}_j = |0\ra\la 0|_{ancilla_0\; j} \otimes |10\ra\la
10|_{clock\; 0,1} + |11\ra\la 11|_{clock\; 0,1}, \quad
j=1,\ldots,n_0,
\]
\[
\Pi^{init \; +}_j = |+\ra\la +|_{ancilla_+\; j} \otimes |10\ra\la
10|_{clock\; 0,1}+ |11\ra\la 11|_{clock\; 0,1}, \quad
j=1,\ldots,n_+.
\]
Here we used the labels {\it input, ancilla${\, }_0$, ancilla${\,
}_+$, clock} to label the subsets of input qubits, ancillas $|0\ra$,
ancillas $|+\ra$, and clock qubits respectively. Also $x_j$ stands
for the $j$-th bit of the string $x$. States invariant under the
projectors above satisfy correct initial conditions.

Introduce non-negative $6$-qubit projectors \bea \Pi^{prop}_j &=&
\frac12 |1\ra\la 1|_{clock\; j-1} \otimes \left(\vphantom{U_j^\dag}
|1\ra\la 1|_{clock\; j} + |0\ra\la 0|_{clock\; j}\right. \nn \\
&& {}\left. + |1\ra\la 0|_{clock\; j} \otimes U_j + |0\ra\la
1|_{clock\; j} \otimes U_j^\dag\right) \otimes
|0\ra\la 0|_{clock\; j+1} \nn \\
&& {} + |000\ra\la 000|_{clock\; j-1,j,j+1} +|111\ra\la
111|_{clock\; j-1,j,j+1}  \eea where $j=1,\ldots,L$. States
invariant under the projectors above obey the correct propagation
rules relating computational states at different time steps.
Therefore we arrive at
\[
\calH=\left\{|\Phi\ra \in (\CC^2)^{\otimes N}\, : \, \Pi^{init\;
x}_j\, |\Phi\ra = \Pi^{init\; 0}_j\, |\Phi\ra = \Pi^{init \; +}_j\,
|\Phi\ra = \Pi^{prop}_j\, |\Phi\ra = |\Phi\ra \quad \mbox{for all
$j$}\right\}.
\]
Now we can define a circuit Hamiltonian \be  H^{(6)}=\sum_{j=1}^n
(I-\Pi^{init\; x}_j) +\sum_{j=1}^{n_0} (I-\Pi^{init\; 0}_j)
+\sum_{j=1}^{n_+} (I-\Pi^{init\; +}_j) + \sum_{j=1}^L (I-
\Pi^{prop}_j). \label{Hclock} \ee

\begin{lemma}\label{lemma:spectral}
The smallest eigenvalue of the circuit Hamiltonian $H^{(6)}$ in Eq.
(\ref{Hclock}) is $0$. The corresponding eigenspace coincides with
$\calH$, see Eq.~(\ref{eq:history}). The second smallest eigenvalue
of $H^{(6)}$ is $\Delta=\Omega(L^{-3})$.
\end{lemma}
\begin{proof} The first part of the lemma follows directly from the
definition of $\calH$. The analysis performed
in~\cite{KSV:computation} shows that the spectrum of $H^{(6)}$ does not
depend on the circuit $U$. Thus one can compute the spectral gap of
$H^{(6)}$ by considering a trivial circuit composed of identity gates,
$U_j=I$. For the trivial circuit one can ignore the witness qubits
 since $H^{(6)}$ does not act on them. Besides, one can
consider only one type of ancillas, say $|0\ra$, because by conjugating
$H^{(6)}$ with unitary Hadamard operators we can convert $|+\ra$ ancillas to
$|0\ra$ ancillas. By similar arguments, we can assume that
$|x\ra=|0^{n}\ra$. Then we apply the result of Lemma 3.11 for $s=1$ in Ref. \cite{ADLLKR:adia} which
shows that Kitaev's circuit Hamiltonian corresponding to a quantum circuit has a spectral
gap $\Delta=\Omega(L^{-3})$.
\end{proof}

\subsubsection{Converting stoquastic and coherent classical verifiers to a stoquastic Hamiltonian}

This section describes the final step in converting a stoquastic or a coherent classical verifier to a stoquastic
Hamiltonian, namely how to represent the final measurement in the circuit. The construction
that we present here is different from the standard one in~\cite{KSV:computation}. The reason for this
modification is that the decision thresholds in \StoqMA{} cannot be amplified and therefore
the standard construction would fail.

Let $V=(n,n_w,n_0,n_+,U)$ be a stoquastic or a coherent classical verifier, where
the circuit $U$ consists of $L$ gates, $U=U_L \cdots U_2\, U_1$.
Define a $3$-qubit non-negative projector
\[
\Pi^{meas}=\Pi_{out}\otimes |10\ra\la 10|_{clock\; L,L+1} +
|00\ra\la 00|_{clock\; L,L+1}.
\]
Here the projector $\Pi_{out}$ corresponds to the final measurement
performed by the verifier $V$, see Def.~\ref{dfn:StoqMA} and Def.~\ref{dfn:ccv}.
 Let $|\Phi\ra\in
\calH$ be a normalized state representing a computational path
starting from a witness state $|\psi\ra\in (\CC^2)^{\otimes n_w}$,
and some input string $x\in \Sigma^n$, see Eq.~(\ref{eq:history}).
One can easily check that \be\label{meas} \la \Phi|
\Pi^{meas}|\Phi\ra = 1 - \frac1{L+1} \left[ 1- \prob(V;\psi,x)
\right]. \ee Thus the subspace $\calH$ contains a state invariant
under $\Pi^{meas}$ iff $\prob(V;\psi,x)=1$ for some witness state
$|\psi\ra$.

Let $H^{(6)}$ be the clock Hamiltonian associated with $V$, see Eq.~(\ref{Hclock}).
 Define a new Hamiltonian
\be\label{perturbation} \tilde{H}=H^{(6)} + \delta\, (I-\Pi^{meas}), \quad
0<\delta\ll \Delta. \ee Let $\lambda(\tilde{H})$ be the smallest
eigenvalue of $\tilde{H}$. Considering $\delta\, (I-\Pi^{meas})$ as
a small perturbation, we can compute $\lambda(\tilde{H})$ as
\[
\lambda(\tilde{H}) = \delta\, \min_{|\phi\ra\in \calH} \la
\phi|(I-\Pi^{meas})|\phi\ra + O(\delta^2).
\]
Taking into account Eq.~(\ref{meas}) one gets
\be\label{tlambda}
\lambda(\tilde{H})=\frac{\delta}{L+1}\, \left(1 -  \max_{\psi}
\prob(V;\psi,x)\right) + O(\delta^2).
\ee
According to Lemma~\ref{lemma:spectral}, $H^{(6)}$ has a spectral gap $\Delta=\Omega(L^{-3})$.
Thus the applicability of the perturbative approach, $\delta \ll
\Delta$, can be ensured by choosing $\delta\ll L^{-3}$.

\subsubsection{Stoquastic $6$-SAT is hard for \MA}
\label{subs:strictproof}
We can define an instance of stoquastic $6$-SAT with a set of projectors
\be\label{calS} \calS=\{ \Pi^{init\; x}_j, \; \Pi^{init\; 0}_j, \;
\Pi^{init\; +}_j, \; \Pi^{prop}_j, \; \Pi^{meas}\}. \ee The total
number of projectors in $\calS$ is $M=n+n_0+n_+ + L+ 1=n^{O(1)}$. If $x$
is yes-instance, then $\prob(V;\psi,x)=1$ for some witness state $|\psi\ra$,
see Lemma~\ref{lemma:ccv},
and thus the set of projectors
$\calS$ has a common invariant state. If $x$ is
no-instance, then for any state $|\Phi\ra\in (\CC^2)^{\otimes N}$ there
exists a projector $\Pi\in \calS$ such that
\[
\la \Phi|(I-\Pi)|\Phi\ra\ge \min{\{1,\delta^{-1}\}}\, \la \Phi|\tilde{H}|\Phi\ra/M \ge
\lambda(\tilde{H})/M,
\]
 and therefore
$\la \Phi|\Pi|\Phi\ra \le 1-\lambda(\tilde{H})/M$.
Taking into account Eq.~(\ref{tlambda}) and the soundness condition
from Lemma~\ref{lemma:ccv} one gets $\la \Phi|\Pi|\Phi\ra \le 1-(2/3)\, \delta
M^{-1} (L+1)^{-1} = 1-n^{-O(1)}$. Thus stoquastic $6$-SAT
defined by Eq.~(\ref{calS}) obeys both the completeness and soundness conditions.

\subsection{Proofs of lemmas in Section~\ref{sec:stoqma}}
\label{sec:proofslem}

\begin{proof}[Proof of Lemma \ref{lemma:Xform}]
By definition, $H=\sum_{S} H_S$ where $H_S$ is a stoquastic Hamiltonian acting on qubits from a
set $S$, $|S|\le k$. By adding the identity factors we can assume
that every term $H_S$ acts on a subset of exactly $k$ qubits.
Applying a shift $H\to H+ \beta\, I$, if necessary, we can assume
that all matrix elements of $H_S$ are non-positive (for all $S$).

Any $k$-qubit Hermitian operator $R$ with non-positive matrix
elements can be written as
\[
R=\frac12 \sum_{x,y\in \Sigma^k} R_{x,y}\, (|x\ra\la y| + |y\ra\la
x|),  \quad R_{x,y}\le 0.
\]
Clearly, for any string $x\in \Sigma^k$ one can construct a quantum
circuit $U_x$ with $X$ gates such that $|x\ra=U\, |0^{k}\ra$.
Analogously, for any pair of strings $x\ne y\in \Sigma^k$ one can
construct a quantum circuit $U_{x,y}$ with $X$ and CNOT gates such
that $|x\ra=U_{x,y}\, |0^{k}\ra$, $|y\ra=U_{x,y}\, |10^{k-1}\ra$.
Thus we get
\[
R=\sum_{x\in \Sigma^k}  R_{x,x}\,  U_x  \left( |0\ra\la 0 |^{\otimes
k}\right)   U_x^\dag + \frac12 \sum_{x\ne y\in \Sigma^k} R_{x,y} \,
U_{x,y} \left( X\otimes |0\ra\la 0|^{k-1}\right)  U_{x,y}^\dag.
\]
Here $U_x$ and $U_{x,y}$ are quantum circuits on $k$ qubits with $X$
and CNOT gates. Applying this decomposition to every term $H_S$
separately, and normalizing the coefficients, we arrive at
Eq.~(\ref{Xform}).\end{proof}

\begin{proof}[Proof of Lemma \ref{lemma:t}]
Let us first prove Eq.~(\ref{ab1}).
The key idea is illustrated in Figure~\ref{fig:Toffoli_trick}.
\begin{figure}[h]
\centerline{ \mbox{\includegraphics[height=1.5cm]{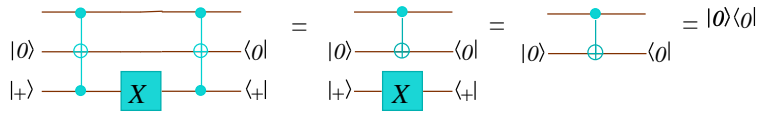}}}
\caption{How to simulate measurement of $|0\ra\la 0|$ by measurement
of $X$.\label{fig:Toffoli_trick}}
\end{figure}
If $k=1$ one can choose $W\, |\psi\ra = T[1,3;2]\, |\psi\ra\otimes
|0\ra \otimes |+\ra$, where $T[1,3;2]$ is the Toffoli gates with
control qubits $1,3$ and target qubit $2$, that is
\[
T[1,3;2]\, |a,b,c\ra = |a,b\oplus ac, c\ra.
\]
One can easily check that
\[
T[1,3;2]\, (I\otimes I \otimes X)\, T[1,3;2]^\dag =
\mbox{CNOT}[1;2]\otimes X.
\]
Accordingly, $W^\dag \, (I\otimes I \otimes X)\, W=\la
0_2|\, \mbox{CNOT}[1;2]\, |0_2\ra \la +|X|+\ra = |0\ra\la 0|$, see
Figure~\ref{fig:Toffoli_trick}. For arbitrary $k$ one can use $k$
copies of the ancilla $|0\ra$ and $k$ Toffoli gates, i.e., $W\,
|\psi\ra = \prod_{j=1}^k T[j,2k+1;j+k]\, |\psi\ra\otimes
|0\ra^{\otimes k} \otimes |+\ra$ (all Toffoli gates in the product
commute). The proof of Eq.~(\ref{ab2}) is the same except for not using
the ancilla $|+\ra$, i.e.,
$W\,
|\psi\ra = \prod_{j=1}^{k-1} T[j,2k-1;j+k-1]\, |\psi\ra\otimes
|0\ra^{\otimes k-1}$.
\end{proof}

\bibliographystyle{hunsrt}

\end{document}